\documentclass[11pt, a4paper]{article}
\usepackage{newtxtext}
\usepackage{newtxmath}

\usepackage[utf8]{inputenc}
\usepackage[T1]{fontenc}
\usepackage{textcomp}
\usepackage{microtype}

\usepackage{geometry}
\usepackage{bm}
\usepackage{physics}

\usepackage{graphicx}
\usepackage{xcolor}

\usepackage{array}
\usepackage{booktabs}
\usepackage{multirow}

\usepackage{siunitx}
\usepackage{authblk}
\usepackage{orcidlink}

\usepackage{tikz}
\usetikzlibrary{arrows.meta, positioning, shapes.geometric}

\usepackage{hyperref}
\usepackage[style=numeric-comp, autocite=superscript, sorting=none]{biblatex}
\begin{document}

\title{A Symmetry-Constrained Fourier--Morse Framework for Compact Anisotropic Interaction Potentials}
\author[1]{Hadis Ghodrati\orcidlink{0009-0004-1451-8086}}
\author[1]{Sibylle Gemming\orcidlink{0000-0003-0455-1945}}
\author[2]{Florian Günther\orcidlink{0000-0001-5002-4172}}
\author[1,3]{Jeffrey Kelling*\orcidlink{0000-0003-1761-2591}}

\affil[1]{Institute of Physics, Technische Universität Chemnitz, 09107 Chemnitz, Germany.}
\affil[2]{Departamento de Física, Universidade Estadual Paulista, Instituto de Geociências e Ciências Exatas, 13506-900 Rio Claro, SP, Brazil.}
\affil[3]{Institute for Radiation Physics, Helmholtz-Zentrum Dresden - Rossendorf, 01328 Dresden, Germany.}

\date{September 14, 2026}

\maketitle

\begin{abstract}

Large-scale coarse-grained simulations of anisotropic particles require compact interaction models that retain orientation-dependent energetics. We present a symmetry-constrained Fourier--Morse framework in which the radial interaction is described by a Morse potential and its orientational dependence by Fourier expansions. The representation converges systematically with harmonic resolution, allows known orientational symmetries to be imposed directly, and supports further reduction through harmonic truncation and coefficient pruning. Its explicit Fourier structure also provides a natural basis for constructing or modifying model interactions with prescribed orientational symmetries. The parameterization requires only a sampled interaction landscape and is therefore independent of the method used to generate the reference data. We demonstrate the approach for four interaction classes of chiral $\alpha$-polyalanine helices, representing more than \num{300000} reference energy values with tens to a few hundred coefficients while reproducing equilibrium interaction features with meV- and m\AA{}-level errors. As a proof of concept, molecular-dynamics simulations using the reduced analytical potentials produce stable low-temperature configurations exhibiting local ordering motifs qualitatively consistent with those identified previously by Monte Carlo simulated annealing.

\end{abstract}
\section{Introduction}

Self-assembled molecular layers provide a versatile route to ordered
two-dimensional interfaces whose properties depend on molecular packing,
orientational order, domain structure, and assembly
dynamics~\autocite{SAM_organic_e_2017,SAM_structure_defects_2013}. Such
systems are relevant to molecular electronics and spintronics, where
interfacial molecular organization can influence charge and spin
transport~\autocite{spintronic_review_2016}. Chiral molecular layers provide
an important example, as their organization can give rise to
chiral-induced spin selectivity (CISS), thereby coupling molecular structure
to spin-dependent electron
transmission~\autocite{CISS_SAM_experiment_2024,CISS_review_2025}.
Understanding and controlling surface self-assembly requires access not only
to favorable local configurations but also to collective behavior such as
long-range ordering, growth kinetics, domain formation, defect evolution,
and the accommodation of substrate-induced
disorder~\autocite{SAM_organic_e_2017,SAM_dynamics_defects_2005,
SAM_simulation_review_2016}. Accessing these length and time scales motivates
large-scale coarse-grained approaches, including molecular dynamics (MD) and
kinetic Monte Carlo (KMC), which require accurate and computationally efficient anisotropic interaction
models~\autocite{SAM_simulation_review_2016,SAM_KMC_2007,Noid_CG_review_2023}.

Coarse-grained interactions can be represented at different levels of
complexity. Multidimensional tabulated potentials provide a direct means of
retaining detailed interaction landscapes generated from electronic-structure
calculations, classical force fields, machine-learning models, or other
reference methods~\autocite{comp_interaction_2025,book_computational_2022},
but their use requires storage and interpolation of multidimensional
interaction data. Conventional analytical pair potentials, including the
Lennard--Jones and Morse forms, are compact and computationally efficient but
cannot alone describe the orientational dependence characteristic of
anisotropic particles~\autocite{Morse_original,LJ_classic}.
Analytical anisotropic models such as the Gay--Berne potential,
patchy-particle interactions, and related orientation-dependent coarse-grained
models encode this dependence through prescribed functional
forms~\autocite{GayBerne_1995,patchy_review_2011,helical_potential_1997,
babadi_CG_2006,Curco_CG_oriented_helical_2007,Sutherland_chiralCG_3D_2019}.
Machine-learning representations provide substantially greater functional
flexibility and can reproduce complex configurational
dependence~\autocite{ML_coarse_grained_MD_2019,ML_CG_2023}, although their
parameterization is not generally expressed directly in terms of physically
interpretable angular harmonics. This motivates a lightweight representation
that combines sufficient flexibility to reconstruct sampled anisotropic
interaction landscapes with a compact analytical form, explicit symmetry
control, and systematic convergence with increasing model complexity.

Here, we develop such a representation by combining a three-parameter Morse
description of the radial interaction with Fourier expansions that introduce
orientational dependence into the Morse parameters. Fourier bases provide a
natural representation of periodic angular variables~\autocite{torsional_fourier_1989,Noid_CG_review_2023}
and allow the angular resolution to be increased systematically through the
retained harmonic orders. Known rotational and interchange symmetries can be
incorporated directly by restricting the admissible Fourier terms, providing
an exact reduction of the parameter space before approximate model reduction
is applied. Harmonic truncation and coefficient pruning then allow the
remaining representation to be reduced in a controlled manner. The resulting
model replaces a densely sampled interaction landscape by a comparatively
small set of analytical coefficients and avoids multidimensional table
interpolation. Because individual Fourier modes correspond directly to characteristic angular periodicities, the same representation also provides a natural basis for constructing or modifying model interactions with prescribed orientational symmetries~\autocite{inverse_SAM_2018,inverse_SAM_2D_2006}.

Chiral and helical particles provide a particularly suitable setting in which to examine these features because their molecular geometry introduces well-defined rotational, screw-related, and interchange symmetries. Their interactions may depend on intermolecular separation, axial rotation angles, handedness, and relative axial direction, while exhibiting periodicities inherited from the molecular screw structure~\autocite{bio_helices_pot_2007,helical_potential_1997,Curco_CG_oriented_helical_2007,Hadis_2026}. We therefore use $\alpha$-polyalanine ($\alpha$PA) helices as a symmetry-rich demonstration system and employ previously reported tabulated pair-interaction landscapes as reference data~\autocite{Hadis_2026}. The representation is not tied to a particular molecular identity or to the method used to generate the reference landscape; the $\alpha$PA system serves here to test the convergence, symmetry reduction, pruning, and practical applicability of the analytical representation.

In this work, we formulate a compact Fourier--Morse representation for
anisotropic pair interactions in a two-dimensional surface geometry and
establish its systematic convergence with harmonic resolution. We then
quantify the additional reduction achievable through symmetry adaptation and
coefficient pruning, with particular emphasis on the screw and interchange
symmetries of chiral helical systems. Using $\alpha$PA as a symmetry-rich
demonstration case, we assess the resulting trade-off between reconstruction
accuracy and model compactness and show that the reduced analytical potential
can be used in proof-of-concept molecular dynamics simulations to produce
stable low-energy arrangements qualitatively consistent with those identified
previously by Monte Carlo simulated annealing~\autocite{Hadis_2026}. The
fitting, symmetry-reduction, pruning, and potential-evaluation workflow is
implemented in the open-source Python package
\texttt{ChiMorse}~\autocite{chimorse}. The general formulation is presented
first, followed by the $\alpha$PA demonstration, the analysis of convergence
and model reduction, and the molecular-dynamics demonstration.

\section{General Framework}
\label{sec:general_framework}

We consider anisotropic pair interactions described by an intermolecular separation and periodic orientational coordinates. The Fourier--Morse representation introduced below provides a compact analytical description of such sampled interaction landscapes.

For the formulation developed here, the interacting particles are rigid chiral helices confined to a common plane. Their relative configuration is specified by the intermolecular separation and two axial rotation angles, while handedness and relative axial direction define distinct interaction classes. Each class is represented by an independent potential unless an exact symmetry relation establishes equivalence between them.

\subsection{Reference Interaction Data and Angular Coordinates}
\label{sec:reference_coordinates}

Consider two rigid helical molecules with geometrically parallel molecular axes. Their in-plane relative configuration is described by the axis-to-axis separation $r$ and the axial rotation angles $\varphi_1$ and $\varphi_2$. The reference interaction considered here therefore has the form
\begin{equation}
E_{\mathrm{ref}}(r,\varphi_1,\varphi_2).
\label{eq:reference_landscape_general}
\end{equation}

All other configurational degrees of freedom, including mutual tilt, relative axial displacement, and internal molecular conformations, are assumed to be fixed or treated through separate interaction models. Extension to additional coordinates is possible in principle but is outside the scope of the present work.

For each sampled angular configuration $(\varphi_1,\varphi_2)$, the reference data should include a radial profile spanning the physically relevant repulsive and attractive regions. This profile provides the information required to determine the local equilibrium separation, well depth, and radial shape of the interaction. The angular sampling should cover either the full periodic domain or a symmetry-equivalent irreducible region with sufficient resolution to capture the relevant angular harmonics.

For the symmetry analysis, the physical rotation angles are transformed to the collective coordinates
\begin{equation}
\chi = \varphi_1 - h\varphi_2,
\qquad
\psi = \varphi_1 + h\varphi_2,
\label{eq:chi_psi_general}
\end{equation}
where $h=+1$ for molecules of equal handedness and $h=-1$ for molecules of opposite handedness.

The coordinate $\chi$ describes the relative angular registry of the molecular pair, whereas $\psi$ describes their joint angular phase. In these coordinates, the screw and molecular-interchange symmetries of helical particles take particularly simple forms, allowing the corresponding periodicity and parity relations to be imposed directly on the Fourier basis. Handedness and relative axial direction determine which symmetry constraints apply and are specified explicitly for the $\alpha$PA demonstration in Sec.~\ref{sec:demonstration}.

\begin{figure}[!htbp]
\centering
\includegraphics[width=0.58\textwidth]{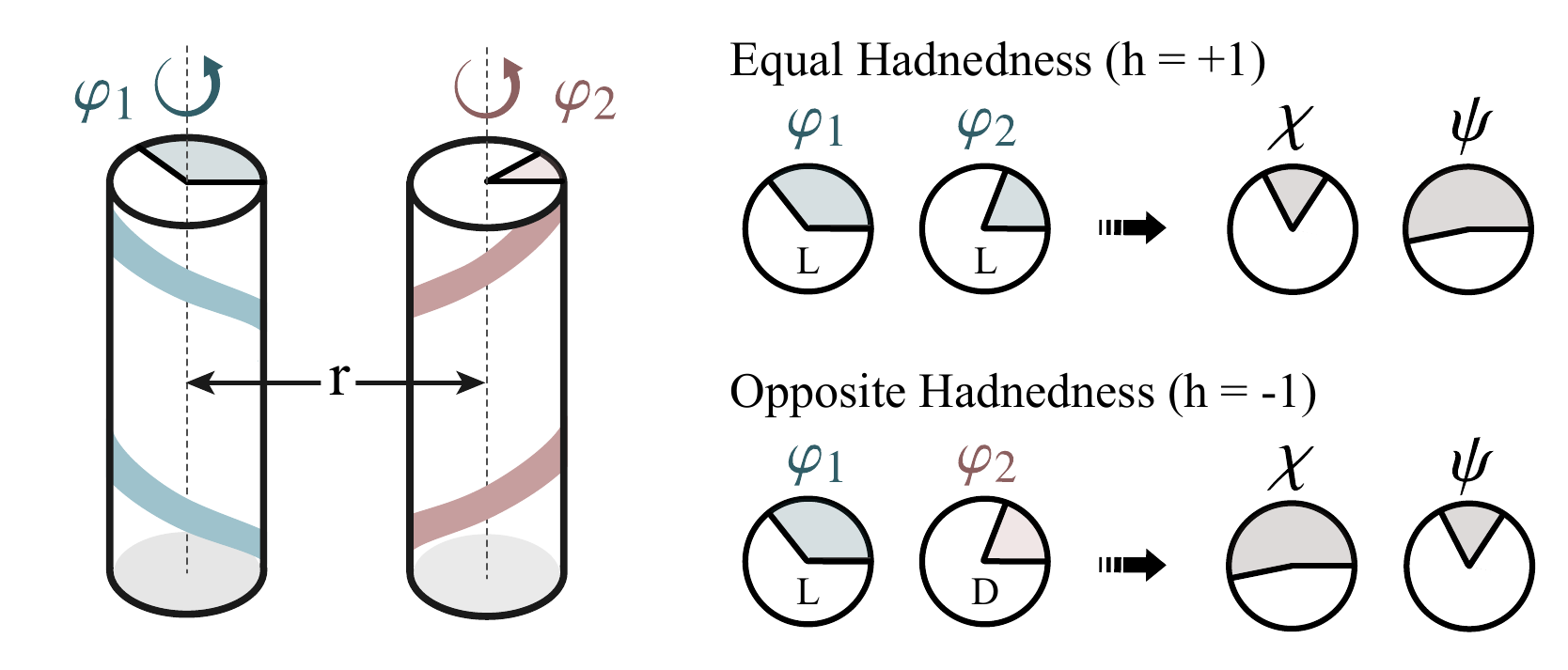}
\caption{
Configurational coordinates for a pair of rigid chiral helical molecules.
The pair is described by the axis-to-axis separation $r$ and the
axial rotations $\varphi_1$ and $\varphi_2$.
The collective coordinates are defined as
$\chi=\varphi_1-h\varphi_2$ and
$\psi=\varphi_1+h\varphi_2$, with $h=+1$ for equal-handed and
$h=-1$ for opposite-handed pairs.
The letters L and D denote the two molecular handednesses.
}
\label{fig:schematic}
\end{figure}

\subsection{Fourier--Morse Representation}
\label{sec:fourier_morse_model}

For each fixed orientational configuration, the radial interaction is represented by a three-parameter Morse potential,
\begin{equation}
E(r;D,r_e,\alpha)
=
D\left[
e^{-2\alpha(r-r_e)}
-
2e^{-\alpha(r-r_e)}
\right],
\label{eq:morse_general}
\end{equation}
where $D$ is the well depth, $r_e$ is the equilibrium separation, and
$\alpha$ controls the radial width and curvature of the interaction
well~\autocite{Morse_original}. The three parameters provide a convenient
separation of the principal radial characteristics: $D$ sets the interaction
strength, $r_e$ locates the minimum, and, for fixed $D$, $\alpha$ controls
the stiffness and effective range of the interaction. In contrast to Lennard--Jones-type forms, which prescribe fixed algebraic
radial exponents, the Morse form provides an adjustable exponential length
scale through $\alpha$, allowing the radial shape and effective range to vary
without increasing the number of radial parameters. This additional shape parameter makes the Morse form
sufficiently flexible to represent variations in the radial profile while
retaining a compact analytical form.

To represent anisotropy, the Morse parameters are allowed to depend on the collective orientational coordinates,
\begin{equation}
D=D(\chi,\psi),
\qquad
r_e=r_e(\chi,\psi),
\qquad
\alpha=\alpha(\chi,\psi).
\label{eq:angular_morse_parameters}
\end{equation}
The resulting interaction potential is
\begin{equation}
E(r,\chi,\psi)
=
D(\chi,\psi)
\left[
e^{-2\alpha(\chi,\psi)\left(r-r_e(\chi,\psi)\right)}
-
2e^{-\alpha(\chi,\psi)\left(r-r_e(\chi,\psi)\right)}
\right].
\label{eq:full_fourier_morse}
\end{equation}

Each angular parameter field
$f\in\{D,r_e,\alpha\}$
is represented by a truncated two-dimensional Fourier expansion,
\begin{equation}
\begin{aligned}
f(\chi,\psi) = a_{00}
&+ \sum_{m=1}^{M}
\left[
a_{m0}\cos(m\chi)
+
\widetilde{a}_{m0}\sin(m\chi)
\right]
\\
&+ \sum_{n=1}^{N}
\left[
a_{0n}\cos(n\psi)
+
\widetilde{a}_{0n}\sin(n\psi)
\right]
\\
&+ \sum_{m=1}^{M}\sum_{n=1}^{N}
\Big[
a_{mn}\cos(m\chi)\cos(n\psi)
+
b_{mn}\cos(m\chi)\sin(n\psi)
\\
&\qquad\qquad\qquad
+
c_{mn}\sin(m\chi)\cos(n\psi)
+
d_{mn}\sin(m\chi)\sin(n\psi)
\Big].
\end{aligned}
\label{eq:fourier_general_framework}
\end{equation}

The retained harmonic orders determine the angular resolution of the representation and provide a systematic route to increasing model complexity. Because the expansion is linear in the Fourier coefficients, the parameter fields can be fitted efficiently and subsequently reduced using the symmetry and pruning procedures described below. The explicit modal structure also provides an intuitive connection between
individual Fourier components and characteristic angular periodicities. This
provides a natural basis for constructing or modifying model interactions
with prescribed rotational symmetries by selecting appropriate Fourier
components.

A reduced form may treat $\alpha$ as angle-independent while retaining the angular dependence of $D(\chi,\psi)$ and $r_e(\chi,\psi)$. This compact model is intended primarily to reproduce the equilibrium interaction landscape with fewer parameters. The full representation retains $\alpha(\chi,\psi)$ and therefore also describes orientational variation in the off-equilibrium radial profiles. The two variants are compared quantitatively for the $\alpha$PA demonstration in Sec.~\ref{sec:pa_parameterization}.

\subsection{Symmetry-Adapted Fourier Basis}
\label{sec:general_symmetry}

The Fourier representation does not require a particular molecular symmetry. When known invariances are present, however, they can be incorporated before fitting by restricting the basis to symmetry-compatible Fourier terms. The applicable constraints depend on the particle symmetry and the chosen orientational coordinates. For the chiral helical systems considered as the demonstration case in this work, two relevant examples are screw-related rotational periodicity and a particle-interchange symmetry.

For a particle pair possessing screw-related rotational periodicity, let $\phi_E$ denote the rotational step associated with an equivalent configuration. The corresponding operation can be written as
\begin{equation}
\varphi_1\rightarrow\varphi_1+\phi_E,
\qquad
\varphi_2\rightarrow\varphi_2+h\phi_E .
\label{eq:screw_phi_transformation}
\end{equation}
With the coordinate convention of Eq.~\eqref{eq:chi_psi_general} and since $h^2=1$, this gives
\begin{equation}
\chi\rightarrow\chi,
\qquad
\psi\rightarrow\psi+2\phi_E .
\label{eq:screw_transformation_general}
\end{equation}
Each angular parameter field $f\in\{D,r_e,\alpha\}$ must therefore satisfy
\begin{equation}
f(\chi,\psi)
=
f(\chi,\psi+2\phi_E).
\label{eq:screw_periodicity_general}
\end{equation}
If $n_0$ is defined by $n_0(2\phi_E)=2\pi$, the admissible harmonics in $\psi$ are restricted to
\begin{equation}
n=\ell n_0,
\qquad
\ell=0,1,\ldots,L.
\label{eq:allowed_screw_harmonics_general}
\end{equation}
The corresponding representation is
\begin{equation}
f(\chi,\psi)
=
A(\chi)
+
\sum_{\ell=1}^{L}
\left[
B_{\ell}(\chi)\cos(\ell n_0\psi)
+
C_{\ell}(\chi)\sin(\ell n_0\psi)
\right],
\label{eq:screw_reduced_fourier}
\end{equation}
where $A(\chi)$, $B_{\ell}(\chi)$, and $C_{\ell}(\chi)$ are Fourier series in $\chi$.

A separate reduction may arise from a particle-interchange operation when that symmetry maps the pair configuration onto an equivalent state within the same interaction class. If its action in the collective coordinates is
\begin{equation}
(\chi,\psi)\rightarrow(-\chi,\psi),
\label{eq:interchange_mapping}
\end{equation}
then
\begin{equation}
f(\chi,\psi)=f(-\chi,\psi),
\label{eq:chi_even_symmetry}
\end{equation}
so that the parameter field is even in $\chi$ and all terms containing $\sin(m\chi)$ are excluded. If the interchange operation does not preserve the interaction class, no such parity restriction is imposed.

The resulting symmetry-adapted basis contains only terms compatible with the
invariances of the interaction class. It therefore provides a reduced basis
for fitting and, conversely, a natural basis for constructing model
interactions with prescribed orientational symmetries.

\subsection{Parameterization and Model Reduction}
\label{sec:general_fitting_reduction}

The construction of the analytical model proceeds in two stages: local extraction of the radial Morse parameters at each orientational configuration, followed by a global Fourier representation of their angular dependence.

For reference radial profiles whose energy approaches zero at large
separation, the well depth $D$ and equilibrium separation $r_e$ at each
sampled configuration $(\chi_i,\psi_j)$ are obtained from the interaction
minimum,
\begin{equation}
D(\chi_i,\psi_j)
=
-\min_{r} E_{\mathrm{ref}}(r,\chi_i,\psi_j),
\qquad
r_e(\chi_i,\psi_j)
=
\operatorname*{argmin}_{r}
E_{\mathrm{ref}}(r,\chi_i,\psi_j).
\label{eq:minimum_general}
\end{equation}
With $D$ and $r_e$ fixed, $\alpha(\chi_i,\psi_j)$ is determined by fitting
the Morse form of Eq.~\eqref{eq:morse_general} to the corresponding radial
profile. When different regions of the radial profile have different physical
relevance, a weighting function may be applied to the residuals during this
fit. This procedure yields discrete angular fields
${D_{ij},r_{e,ij},\alpha_{ij}}$ on the sampled orientational grid.

The angular dependence of each parameter field
$f\in\{D,r_e,\alpha\}$ is subsequently fitted using the applicable
symmetry-adapted Fourier basis. For fixed harmonic orders, the representation
is linear in the Fourier coefficients, which are obtained by linear
least-squares fitting. Symmetry restrictions are therefore incorporated in
the basis before parameter estimation rather than imposed on the fitted
coefficients afterward.

After parameterization, the complexity of the Fourier representation is reduced in two steps.
First, convergence with angular resolution is assessed by systematically
increasing the retained Fourier harmonics. Second, within a chosen harmonic basis, further reduction is obtained
for each parameter field by ranking its fitted coefficients by magnitude,
progressively removing the weakest terms, and refitting the remaining
coefficients after each reduction step.

Reconstruction accuracy is quantified by the root-mean-square error (RMSE)
between the reference and fitted parameter fields over the sampled angular
configurations. The resulting convergence and pruning behavior is used to
select a representation that balances reconstruction accuracy and model
compactness for the interaction landscape under consideration.


\subsection{Potential Evaluation and Implementation}
\label{sec:chimorse_implementation}

Once the Fourier coefficients have been determined, the interaction energy is
evaluated directly from the analytical representation. Radial forces and
orientational torques are obtained as analytical derivatives of the fitted
Fourier--Morse energy representation with respect to the corresponding
configurational coordinates. The resulting potential therefore requires only
the fitted coefficients and avoids multidimensional interpolation of either
the interaction energy or its derivatives.

\begin{figure}[!htbp]
\centering
\includegraphics[width=\textwidth]{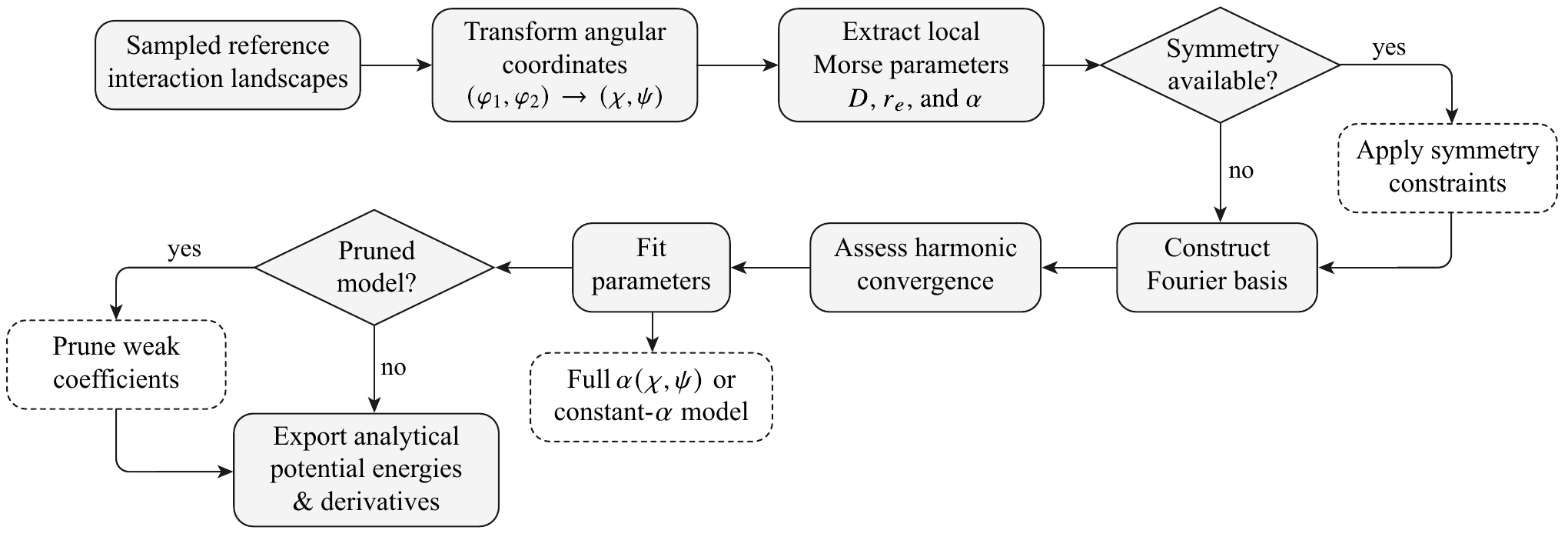}
\caption{
Workflow for constructing the analytical interaction model from sampled
reference data. Local Morse parameters are represented in a Fourier basis,
with symmetry constraints incorporated when applicable. The angular resolution
is selected from the convergence of the reconstruction RMSE with increasing
harmonic order, followed where appropriate by coefficient pruning to reduce
model complexity before molecular simulation.
}
\label{fig:chimorse_workflow}
\end{figure}

The complete construction procedure is summarized in
Fig.~\ref{fig:chimorse_workflow}. Starting from sampled reference interaction
data, local Morse parameters are extracted and their angular dependence is
represented in a symmetry-adapted Fourier basis. The required angular
resolution is determined by monitoring the reconstruction RMSE as the retained
Fourier harmonic orders are increased. Once the harmonic basis has been
selected, coefficient pruning may be applied to obtain a more compact
representation. Depending on the required description of the radial profiles,
either the full model with angle-dependent $\alpha(\chi,\psi)$ or the reduced
constant-$\alpha$ form may be retained for subsequent simulation.

The parameterization, symmetry adaptation, convergence analysis, and pruning workflow is implemented in the open-source Python package \texttt{ChiMorse}~\autocite{chimorse}. Fitted coefficients can be exported for use in external simulation codes, where energies, forces, and torques are evaluated directly from the analytical expressions. Software-specific details and data formats are provided with the package documentation.

\section{Results and Discussion: Demonstration for
\texorpdfstring{$\alpha$}{alpha}-Polyalanine Helices}
\label{sec:demonstration}

\subsection{Reference System and Interaction Classes}
\label{sec:pa_reference_classes}

The representation is demonstrated using previously reported tabulated
pair-interaction landscapes for rigid $\alpha$-polyalanine ($\alpha$PA)
helices derived from ab initio-based calculations~\autocite{Hadis_2026}.
These data provide a symmetry-rich reference system for assessing the
parameterization and model-reduction procedure. Only the sampled interaction
landscape enters the present parameterization; its construction is therefore
not tied to the particular method used to generate these reference data.

The $\alpha$PA helices are treated as rigid bodies with geometrically parallel
molecular axes. Their relative configuration is described by the axis-to-axis
separation $r$, the axial rotations $\varphi_1$ and $\varphi_2$, and the
relative axial displacement $\zeta$. For the surface-supported geometry
considered here, the molecular centers are constrained to the same axial
position, so that
\begin{equation}
E_{\mathrm{ref}}
\left(r,\varphi_1,\varphi_2\right)
=
E_{\mathrm{ref}}
\left(r,\varphi_1,\varphi_2,\zeta=0\right).
\label{eq:pa_reference_landscape_reduced}
\end{equation}
This subset defines the pair-interaction landscapes used in the present
analysis. The reference data were generated on a reduced angular domain and
extended over the full periodic range using the known molecular screw
symmetry. Details of the angular sampling, symmetry completion, and radial
grid are provided alongside the sampled data~\autocite{data_PA}.

In addition to these continuous coordinates, the interaction depends on the
relative handedness and axial direction of the two helices. Here,
\textit{parallel} and \textit{antiparallel} refer to the relative
N$\rightarrow$C directions; the molecular axes remain geometrically parallel
in both cases. Combining equal or opposite handedness with parallel or
antiparallel N$\rightarrow$C alignment gives the four interaction classes
summarized in Table~\ref{tab:pa_interaction_classes}. Each class is represented
by a separate interaction landscape, with symmetry constraints applied where
an operation maps configurations onto equivalent states within that class.

\begin{table}[!htbp]
\centering
\caption{
Interaction classes for the $\alpha$PA demonstration. The parameter $h$
defines the collective coordinates in Eq.~\eqref{eq:chi_psi_general}. The
final column indicates whether an interchange symmetry operation maps the
pair onto an equivalent configuration within the same interaction class.
}
\label{tab:pa_interaction_classes}
\begin{tabular}{cccccc}
\toprule
{}
& Class
& Relative handedness
& Axial alignment
& $h$
& Interchange symmetry \\
\midrule
    \begin{minipage}{.08\textwidth}
      \includegraphics[width=\linewidth]{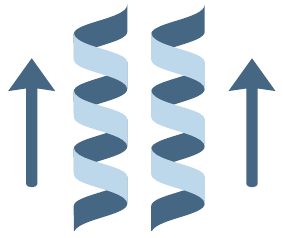}
    \end{minipage}
    & EP & Equal    & Parallel     & $+1$ & yes \\
    \begin{minipage}{.08\textwidth}
      \includegraphics[width=\linewidth]{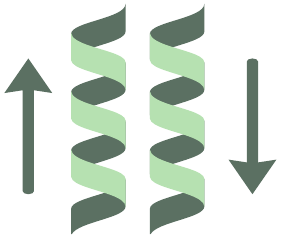}
    \end{minipage}
    & EA & Equal    & Antiparallel & $+1$ & no \\
    \begin{minipage}{.08\textwidth}
      \includegraphics[width=\linewidth]{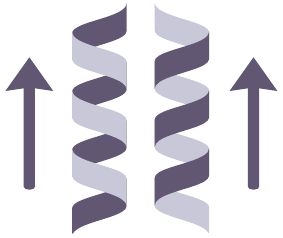}
    \end{minipage}
    & OP & Opposite & Parallel     & $-1$ & yes \\
    \begin{minipage}{.08\textwidth}
      \includegraphics[width=\linewidth]{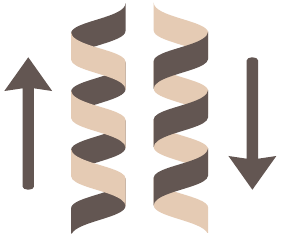}
    \end{minipage}
    & OA & Opposite & Antiparallel & $-1$ & no \\
\bottomrule
\end{tabular}
\end{table}

\subsection{Coordinate and Symmetry Specification}
\label{sec:pa_symmetry}

For the equal-handed classes EP and EA, $h=+1$, whereas for the
opposite-handed classes OP and OA, $h=-1$ in
Eq.~\eqref{eq:chi_psi_general}. The resulting collective coordinates provide
a common representation of the four interaction classes, with $\chi$
describing the relative angular registry and $\psi$ the joint helical phase.

For the $\alpha$PA reference system, the relevant screw-related angular step
is $\phi_E=20^\circ$~\autocite{Hadis_2026}. According to the transformation
derived in Sec.~\ref{sec:general_symmetry}, the corresponding pair operation
leaves $\chi$ unchanged and advances $\psi$ by $40^\circ$. Each angular Morse
parameter therefore satisfies
\begin{equation}
f(\chi,\psi)
=
f\left(\chi,\psi+40^\circ\right),
\qquad
f\in\{D,r_e,\alpha\},
\label{eq:pa_parameter_periodicity}
\end{equation}
giving the fundamental harmonic $n_0=9$. The admissible Fourier modes in
$\psi$ are consequently restricted to
\begin{equation}
n=9\ell,
\qquad
\ell=0,1,\ldots,L.
\label{eq:pa_allowed_psi_modes}
\end{equation}
All symmetry-incompatible $\psi$ harmonics are thus excluded before fitting.

An additional restriction arises from the interchange symmetry operation for
the parallel classes. For EP, the two helices are identical, and simple
interchange of the molecular labels,
$\varphi_1\leftrightarrow\varphi_2$, leaves the physical pair unchanged. In
the collective coordinates this operation leaves $\psi$ invariant and maps
$\chi\rightarrow-\chi$. For OP, the two helices are mirror partners, so a
simple label exchange alone does not generate an equivalent configuration.
The corresponding symmetry operation combines interchange with reflection of
the azimuthal orientations and again acts as
\begin{equation}
(\chi,\psi)\rightarrow(-\chi,\psi).
\label{eq:pa_parallel_interchange}
\end{equation}
A geometrical derivation of the OP transformation is given in the supplementary material.

The parameter fields for both parallel classes are therefore even in $\chi$,
and all Fourier terms containing $\sin(m\chi)$ are excluded. For the
antiparallel classes EA and OA, no equivalent interchange operation acts
within the same interaction class under the adopted geometry, and no parity
restriction in $\chi$ is imposed. Both sine and cosine harmonics are therefore
retained.

The four interaction classes thus illustrate how system-specific molecular
symmetries translate directly into restrictions of the Fourier basis before
harmonic truncation or coefficient pruning. The same correspondence could, in principle, provide a basis for constructing model interactions with prescribed orientational symmetries.

\subsection{Structure of the Reference Interaction Landscapes}
\label{sec:pa_reference_properties}

\begin{figure}[!htbp]
\centering
\includegraphics[width=0.72\textwidth]{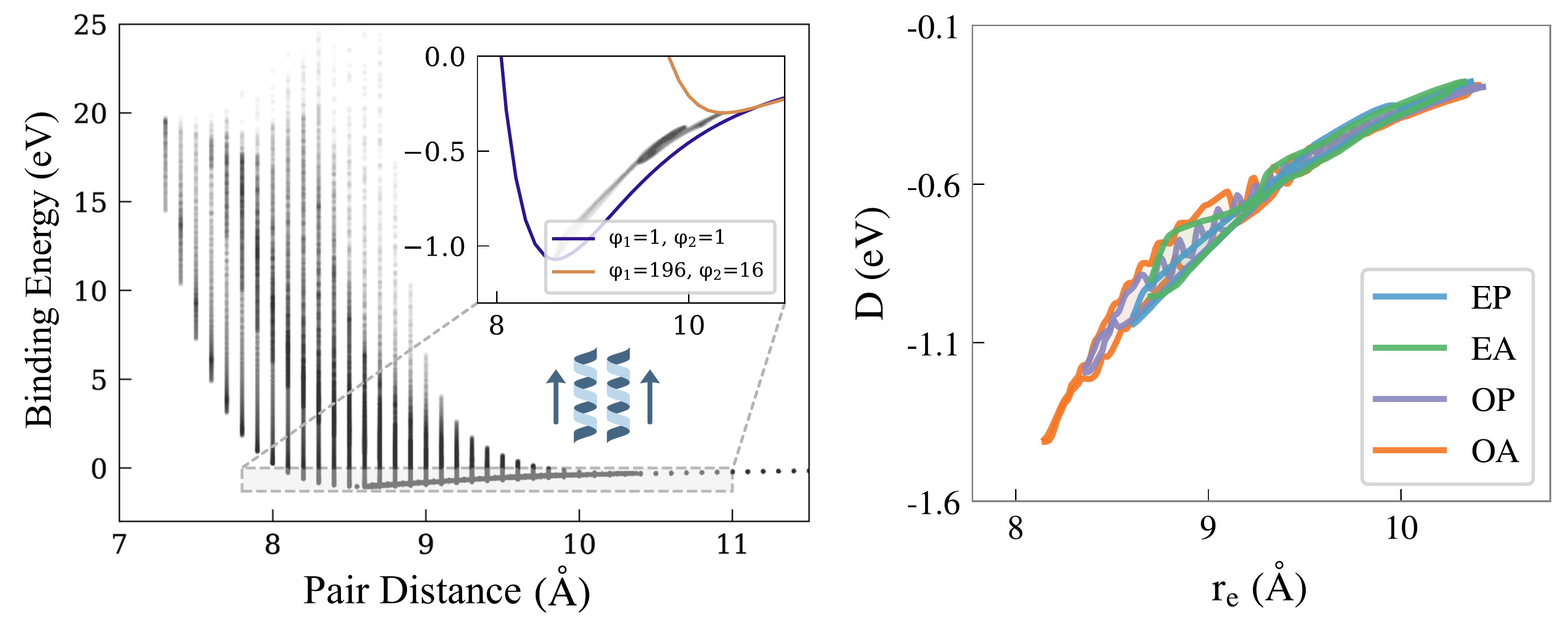}
\caption{
Radial characteristics of the $\alpha$PA reference interaction landscapes.
Left: interaction-energy profiles across the angular configurations represented
for the EP class; the inset shows the corresponding variation in the well depth
$D$ and equilibrium separation $r_e$. Right: extracted $(r_e,D)$ values for
the EP, EA, OP, and OA interaction classes.
}
\label{fig:D_re_3}
\end{figure}

We first examine the radial and angular structure of the $\alpha$PA reference
interaction landscapes to identify the features that must be captured by the
analytical representation. Figure~\ref{fig:D_re_3} summarizes their radial
characteristics. The left panel shows the interaction-energy profiles of the
EP class across the angular configurations represented in the reference
landscape, while the right panel compares the extracted equilibrium
separations and well depths for all four interaction classes.

The radial profiles exhibit pronounced orientational variation in both the
position and depth of the interaction minimum. The four interaction classes
also occupy distinct regions of the $(r_e,D)$ parameter space, demonstrating
that relative handedness and axial direction influence both the preferred
separation and interaction strength. These differences support the separate
parameterization of the four interaction classes introduced above.

\begin{figure}[!htbp]
\centering
\includegraphics[width=\textwidth]{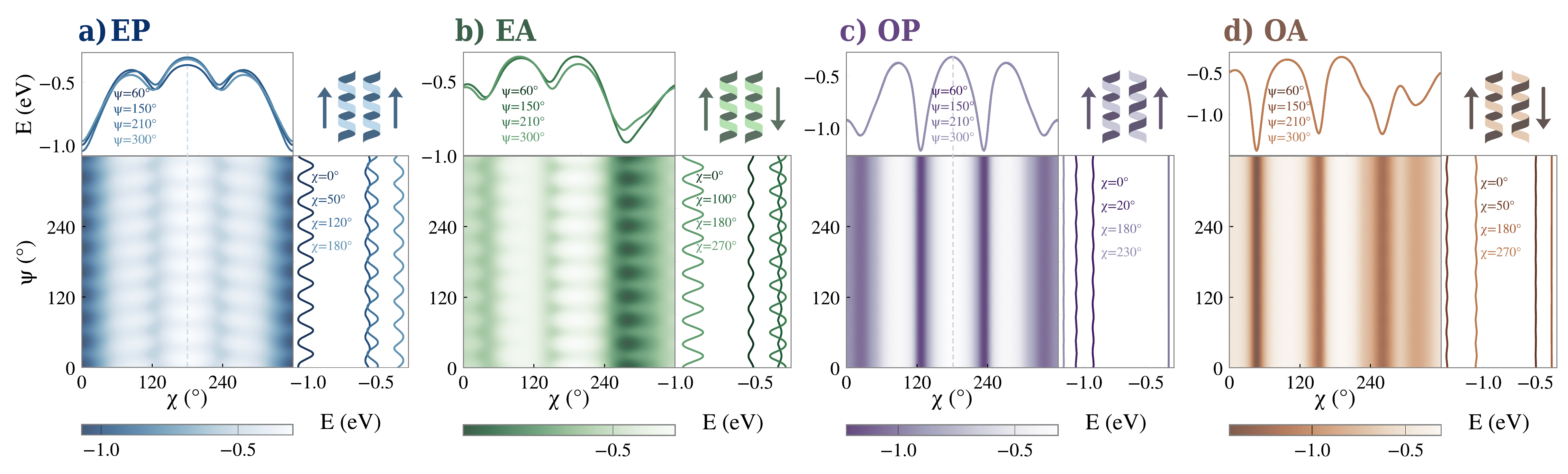}
\caption{
Angular dependence of the reference well depth $D(\chi,\psi)$ for the four
$\alpha$PA interaction classes: (a) EP, (b) EA, (c) OP, and (d) OA. For each
class, the symmetry-completed two-dimensional landscape is shown together with
representative cuts along $\chi$ and $\psi$. For the parallel classes EP and
OP, dashed curves indicate cuts related by the interchange symmetry operation.
}
\label{fig:chi_psi_structure}
\end{figure}

The corresponding angular structure is illustrated in
Fig.~\ref{fig:chi_psi_structure} through the well-depth fields
$D(\chi,\psi)$. For each interaction class, the symmetry-completed
two-dimensional landscape is shown together with representative cuts along the
relative-registry coordinate $\chi$ and the joint angular phase $\psi$.

By construction, the reference landscapes possess the $40^\circ$ screw
periodicity in $\psi$ discussed in Sec.~\ref{sec:pa_symmetry}. Within this
symmetry-constrained structure, the variation along $\chi$ is substantially
stronger than that along $\psi$. The dominant angular corrugation is therefore
associated with the relative registry of the helices, whereas the joint-phase
coordinate provides a weaker screw-periodic modulation. For the parallel
classes EP and OP, the corresponding cuts are consistent with the expected
even dependence on $\chi$, while no analogous parity constraint applies to EA
and OA. The class-dependent variation along $\chi$ further suggests that the
harmonic resolution required to represent the landscapes may differ among the
four interaction classes. These features motivate the harmonic-resolution and model-reduction analysis presented in
Sec.~\ref{sec:pa_parameterization}. The corresponding
$r_e(\chi,\psi)$ landscapes, provided in the supplementary material, exhibit
the same qualitative symmetry structure.

\subsection{Model Parameterization and Reduction}
\label{sec:pa_parameterization}

The four reference landscapes were parameterized using the procedure of
Sec.~\ref{sec:general_fitting_reduction} with the symmetry-adapted bases
specified in Sec.~\ref{sec:pa_symmetry}. Both the compact model with
angle-independent $\alpha$ and the full model with
$\alpha(\chi,\psi)$ were considered. 

\begin{figure}[!htbp]
\centering
\includegraphics[width=\textwidth]{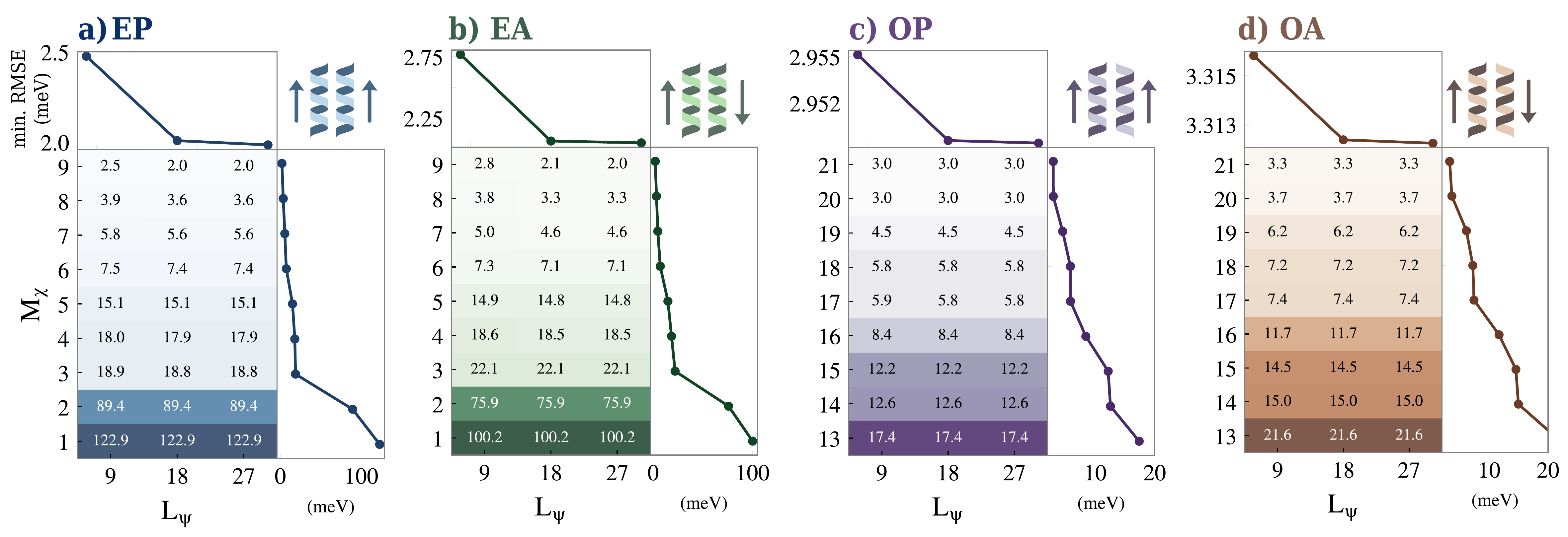}
\caption{
Reconstruction RMSE of the symmetry-adapted Fourier representation of
$D(\chi,\psi)$ as a function of the maximum retained $\chi$ harmonic $M$
and the $\psi$-harmonic family index $L$ for the EP, EA, OP, and OA
interaction classes. Screw symmetry restricts the $\psi$ harmonics to
$n=\ell n_0$ with $n_0=9$. The selected cutoffs correspond to the onset of
the RMSE plateau.
}
\label{fig:pa_harmonic_convergence}
\end{figure}

The required angular resolution was determined from the convergence of the
reconstruction RMSE as the retained Fourier harmonics were increased in
$\chi$ and $\psi$. Figure~\ref{fig:pa_harmonic_convergence} shows this
behavior for the well-depth field $D(\chi,\psi)$. In the $\psi$ direction,
the first symmetry-allowed harmonic, $n=n_0=9$ ($L=1$), captures the dominant
joint-phase dependence for all four interaction classes, with little further
reduction in RMSE from higher harmonics. The convergence in $\chi$ is more
class dependent: EP and EA are adequately represented with $M=8$, whereas
the more structured OP and OA landscapes require $M=20$. The corresponding
analysis for $r_e(\chi,\psi)$ shows the same qualitative behavior and is
provided in the supplementary material.

For the $\alpha$PA data, the local $\alpha$ values were obtained using a
shifted continuous-Poisson weighting with $\lambda=0.8$ in the radial fits.
This asymmetric weighting reduces the influence of the steep short-range
repulsive branch while retaining the low-energy and attractive regions of
the sampled radial profiles. The explicit form of the weighting function and
its implementation are provided in the supplementary material and in the
\texttt{ChiMorse} package.

The angular dependence of $\alpha$ requires somewhat higher resolution for
the equal-handed classes. For EP and EA, the RMSE approaches its plateau near
$L=3$ ($n=27$), whereas for OP and OA the first allowed mode
($L=1$, $n=9$) is sufficient. The required resolution in $\chi$ remains
comparable to that obtained for $D$ and $r_e$. The corresponding convergence
plots for $\alpha(\chi,\psi)$ are given in the supplementary material.

\begin{figure}[!htbp]
\centering
\includegraphics[width=\textwidth]{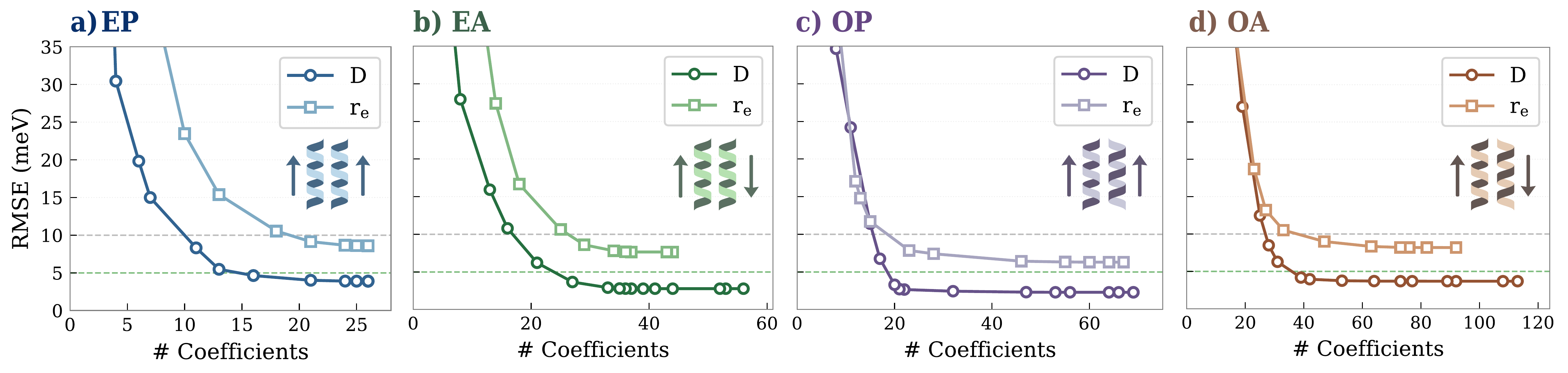}
\caption{
Reconstruction RMSE as a function of the number of retained symmetry-allowed
coefficients during magnitude-based pruning of the compact model. The retained
representations are selected near the onset of the RMSE plateau. Results for
the full model are provided in the supplementary material.
}
\label{fig:pa_pruning_convergence}
\end{figure}

After selection of the harmonic basis, the remaining coefficients were
further reduced by magnitude-based pruning. Figure~\ref{fig:pa_pruning_convergence}
shows the reconstruction RMSE as a function of the number of retained
coefficients for the compact model. Once the dominant coefficients are
included, the RMSE approaches a plateau, indicating that many of the
remaining symmetry-allowed terms contribute only weakly to the reconstructed
landscape. The final coefficient sets were therefore chosen near the onset of
this plateau rather than by applying a universal magnitude threshold. The
full model exhibits the same qualitative pruning behavior, although the
fraction of coefficients retained is class and parameter dependent; the
corresponding analysis is provided in the supplementary material.

Overall, symmetry adaptation, harmonic selection, and coefficient pruning
provide successive reductions in model complexity. The pruning step further
reduces the number of retained coefficients with little change in the
reconstruction RMSE. Table~\ref{tab:error_analysis} summarizes the
reconstruction accuracy of the unpruned symmetry-adapted representations
together with the corresponding model sizes before and after pruning.
Detailed pruning behavior and the reconstruction errors of the pruned
representations are provided in the supplementary material.

\subsection{Reconstruction Accuracy Across Energy Scales}
\label{sec:pa_validation}

We compare the reconstruction accuracy of the compact constant-$\alpha$
and full angle-dependent-$\alpha(\chi,\psi)$ Fourier--Morse
representations. The compact model retains the orientational dependence of
$D(\chi,\psi)$ and $r_e(\chi,\psi)$ while using a single radial-shape
parameter $\alpha$, whereas the full model additionally represents
$\alpha(\chi,\psi)$ as an angular field. Table~\ref{tab:error_analysis}
summarizes the corresponding model sizes and reconstruction errors.

For the parameter fields defining the interaction minima, both
representations provide comparable accuracy. In the full model, the RMSE of
$D(\chi,\psi)$ ranges from $2.97$ to $3.69$~meV and that of
$r_e(\chi,\psi)$ from $5.63$ to $7.51$~m\AA{}. The corresponding compact-model
errors are $2.97$--$3.89$~meV and $6.49$--$7.81$~m\AA{}, respectively.
Thus, both representations accurately reproduce the angular variation of
the equilibrium interaction landscape.

To characterize the energy reconstruction on physically motivated scales,
we define
\begin{equation}
\Delta E_{\rm ref}
=E_{\rm ref}(r,\chi,\psi)-
E_{\min,\rm ref}(\chi,\psi),
\end{equation}
and evaluate the energy RMSE for configurations satisfying
$\Delta E_{\rm ref}\leq2k_{\rm B}T$ and
$\Delta E_{\rm ref}\leq4k_{\rm B}T$. At $T=300$~K,
$k_{\rm B}T=25.85$~meV, so these windows extend approximately
$51.7$ and $103.4$~meV above the corresponding local interaction minimum.
The energy-based definition avoids introducing an arbitrary fixed radial
interval around $r_e$ and provides a common criterion across angular
configurations with different well depths and curvatures.

Within $2k_{\rm B}T$, the energy RMSE is $3.26$--$3.93$~meV for the full
model and $3.69$--$5.04$~meV for the compact model. Expanding the window to
$4k_{\rm B}T$ gives RMSE values of $4.36$--$4.82$~meV and
$5.53$--$7.90$~meV, respectively. Both representations therefore remain
well below the thermal energy scale $k_{\rm B}T$ throughout the
near-equilibrium region considered here. The relatively small difference
between the two models in these windows confirms that the constant-$\alpha$
representation provides an effective compact description when the
near-equilibrium interaction landscape is of primary interest.

The advantage of the angle-dependent $\alpha(\chi,\psi)$ becomes more
pronounced away from the interaction minima. To assess the reconstruction
over a broader portion of the radial profiles, we additionally consider
configurations with $E_{\rm ref}\leq5$~eV. This cutoff retains a wide range
of off-equilibrium configurations while excluding the extreme short-range
repulsive region, which is not relevant to the intended low-energy
applications. Within this region, the full-model RMSE ranges from
$97.88$ to $169.62$~meV, compared with $282.15$--$487.85$~meV for the
compact model. The additional angular dependence of $\alpha$ therefore
substantially improves the reconstruction of the off-equilibrium radial
profiles, consistent with its role in controlling the width and curvature
of the Morse interaction.

These results establish a direct trade-off between model compactness and
radial fidelity. The constant-$\alpha$ representation requires fewer
coefficients and already provides meV-scale accuracy in the thermally
relevant vicinity of the interaction minima. Retaining
$\alpha(\chi,\psi)$ increases the model size but provides substantially
better accuracy over a wider range of radial configurations. The appropriate
representation can therefore be selected according to the energy range and
accuracy required by the intended application.

\begin{table}[!htbp]
\centering
\small
\renewcommand{\arraystretch}{1.15}
\setlength{\tabcolsep}{5.5pt}

\begin{tabular}{@{}llcccc@{}}
\toprule
Model & Quantity & EP & EA & OP & OA \\

\midrule
\multicolumn{6}{l}{\textit{Reference data}} \\

& Reference energy values
& \num{303210} & \num{305607} & \num{321900} & \num{323963} \\

\midrule
\multicolumn{6}{l}{\textit{Full model: angle-dependent $\alpha(\chi,\psi)$}} \\

& Harmonics $(M,L)$
& $(8,3)$ & $(8,3)$ & $(20,1)$ & $(20,1)$ \\

& Fourier coefficients before symmetry reduction
& 2805 & 2805 & 2337 & 2337 \\

& Symmetry-adapted coefficients
& 189 & 357 & 189 & 369 \\

& Coefficients after pruning
& \textbf{115} & \textbf{120} & \textbf{80} & \textbf{160} \\

& RMSE $(D)$ (meV)
& 3.60 & 3.29 & 2.97 & 3.69 \\

& RMSE $(r_e)$ (m\AA{})
& 5.89 & 5.63 & 6.49 & 7.51 \\

& RMSE $(E)$, $\Delta E_{\rm ref}\leq2k_{\rm B}T$ (meV)
& 3.77 & 3.43 & 3.26 & 3.93 \\

& RMSE $(E)$, $\Delta E_{\rm ref}\leq4k_{\rm B}T$ (meV)
& 4.76 & 4.38 & 4.36 & 4.82 \\

& RMSE $(E)$, $E_{\rm ref}\leq5$~eV (meV)
& 97.88 & 143.64 & 169.62 & 140.16 \\

\midrule
\multicolumn{6}{l}{\textit{Compact model: constant $\alpha$}} \\

& Harmonics $(M,L)$
& $(8,1)$ & $(8,1)$ & $(20,1)$ & $(20,1)$ \\

& Fourier coefficients before symmetry reduction
& 646 & 646 & 1558 & 1558 \\

& Symmetry-adapted coefficients
& 54 & 102 & 126 & 246 \\

& Coefficients after pruning
& \textbf{46} & \textbf{65} & \textbf{60} & \textbf{112} \\

& RMSE $(D)$ (meV)
& 3.89 & 3.77 & 2.97 & 3.69 \\

& RMSE $(r_e)$ (m\AA{})
& 7.81 & 7.32 & 6.49 & 7.51 \\

& RMSE $(E)$, $\Delta E_{\rm ref}\leq2k_{\rm B}T$ (meV)
& 4.27 & 4.28 & 3.69 & 5.04 \\

& RMSE $(E)$, $\Delta E_{\rm ref}\leq4k_{\rm B}T$ (meV)
& 5.53 & 5.63 & 5.69 & 7.90 \\

& RMSE $(E)$, $E_{\rm ref}\leq5$~eV (meV)
& 282.15 & 366.33 & 433.08 & 487.85 \\

\bottomrule
\end{tabular}

\caption{
Model complexity and reconstruction accuracy for the full
angle-dependent-$\alpha(\chi,\psi)$ and compact constant-$\alpha$
Fourier--Morse representations of the four $\alpha$PA interaction classes.
Energy errors are evaluated relative to the local reference minimum,
$\Delta E_{\rm ref}=E_{\rm ref}-E_{\min,\rm ref}$.
The first two energy windows correspond to configurations within
$2k_{\rm B}T$ and $4k_{\rm B}T$ of the local minimum at $T=300$~K,
where $k_{\rm B}T=25.85$~meV. The $E_{\rm ref}\leq5$~eV criterion
provides a broader comparison while excluding the strongly repulsive
high-energy portion of the sampled profiles. 
RMSE values reported in the table correspond to the unpruned
symmetry-adapted representations and should therefore be distinguished from
the post-pruning coefficient counts. The numbers of retained coefficients
after pruning are included to indicate the size of the corresponding reduced
representations; detailed pruning results and reconstruction errors of the
pruned models are provided in the supplementary material.
}
\label{tab:error_analysis}
\end{table}


\subsection{Proof-of-Concept Molecular-Dynamics Demonstration}
\label{sec:pa_md_validation}

To assess the applicability of the reduced analytical potentials in
dynamical simulations, their implementation was first tested under
microcanonical conditions. NVE simulations for representative compositions
spanning all four interaction classes show conservation of the total energy
without systematic drift over the simulated trajectories. Representative
energy-conservation tests and simulation details are provided in the
supplementary material.

\begin{figure}[!htbp]
\centering
\includegraphics[width=.8\textwidth]{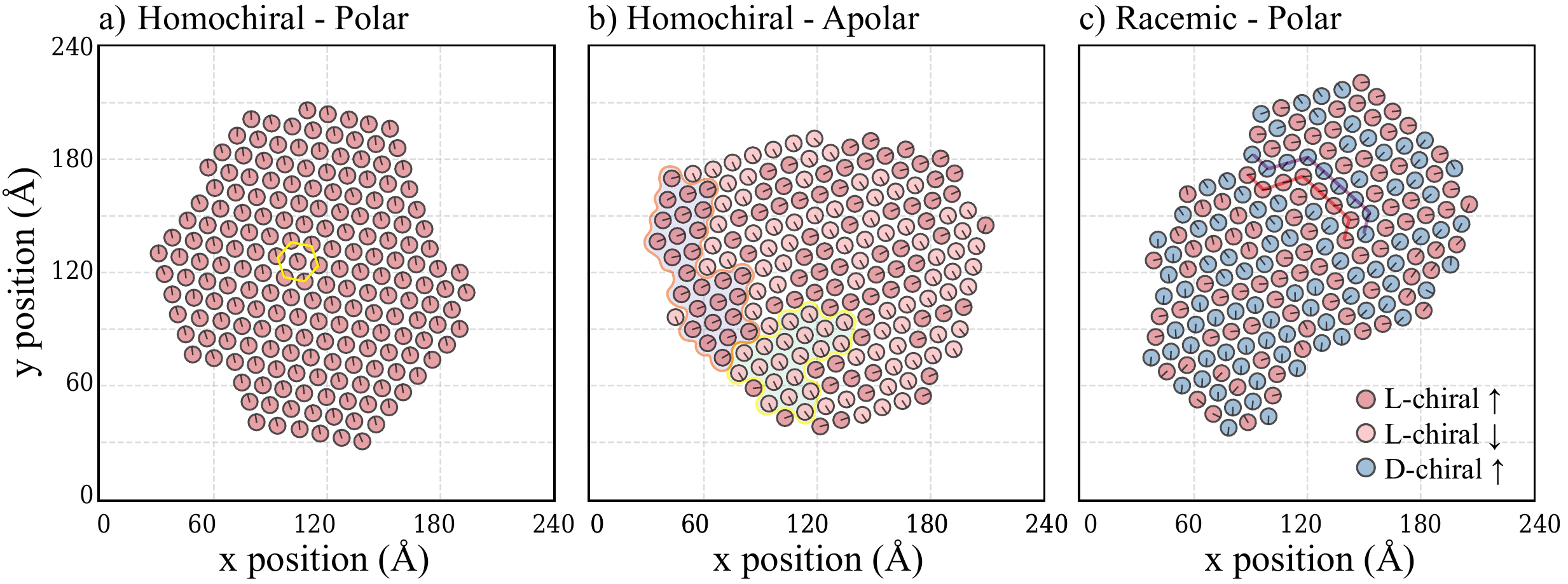}
\caption{
Representative low-temperature configurations obtained from
molecular-dynamics annealing simulations for (a) homochiral polar,
(b) homochiral apolar, and (c) racemic polar systems. Particle colors
distinguish handedness and axial direction according to the graphical key,
while the line within each particle indicates its molecular rotational
orientation. The reduced temperature is defined as
$T^*=k_{\mathrm B}T/E_0$ with $E_0=0.4$~eV. The configurations shown were
obtained at $T^*=0.01$ after stepwise cooling in increments
$\Delta T^*=0.01$, with a nominal cooling rate
$\Delta T^*/t^*_{\mathrm{stage}}=10^{-4}$ per reduced time unit; selected
stages near the structural-relaxation region were extended to
$t^*_{\mathrm{stage}}=1000$. The configurations represent dynamically
accessible low-temperature states rather than optimized minimum-energy
structures.
}
\label{fig:md_structures}
\end{figure}

The reduced potentials were then used in many-particle molecular-dynamics
simulations in which initially disordered configurations were progressively
cooled to low temperature. Three representative systems were considered:
a homochiral polar system containing only EP interactions, a homochiral
apolar system containing EP and EA interactions, and a racemic polar system
containing EP and OP interactions. Representative low-temperature
configurations are shown in Fig.~\ref{fig:md_structures}. The homochiral
polar system forms a compact, approximately hexagonally packed aggregate
with substantial local orientational alignment. The homochiral apolar system
shows local segregation according to axial direction, while the racemic
polar system develops locally ordered rows enriched in particles of the same
handedness.

The observed local ordering tendencies are qualitatively consistent with
motifs identified previously from the tabulated interactions by Monte Carlo
simulated annealing~\autocite{Hadis_2026}. The previous optimization produced
hexagonally packed orientational order in the homochiral polar system,
pronounced segregation of opposite axial directions in the homochiral apolar
system, and handedness-dependent rows in the racemic polar system. The
molecular-dynamics simulations recover the corresponding tendencies locally,
but not necessarily the same long-range structures.

A direct one-to-one comparison of the final states is not expected because
the two procedures serve different purposes. The previous Monte Carlo
annealing was used to minimize the interaction energy toward a
zero-temperature ground-state structure and employed non-dynamical moves,
including particle exchanges, that facilitate large configurational
rearrangements. In contrast, molecular dynamics follows continuous
trajectories at finite temperature and therefore retains thermal and entropic
effects. As the system is cooled, transitions between metastable
configurations become progressively less probable on the accessible
simulation timescale, so relaxation toward the global minimum may require
substantially longer trajectories than those considered here. The present
simulations therefore provide a proof of concept that the reduced analytical
potentials support stable many-particle dynamics while retaining the principal
local structural preferences encoded in the reference interactions.

\section{Conclusion and Outlook}
\label{sec:conclusion}

We have developed a symmetry-constrained Fourier--Morse framework for
constructing compact analytical pair potentials from sampled anisotropic
interaction landscapes. The radial dependence is described by a
three-parameter Morse form, while the orientational dependence of the Morse
parameters is represented through Fourier expansions. When known invariances
are available, symmetry-incompatible Fourier terms are excluded before
fitting, providing an exact reduction of the parameter space. Convergence of
the reconstruction error with harmonic resolution and subsequent coefficient
pruning then provide controlled numerical reduction of the remaining
representation.

The approach was demonstrated using previously reported tabulated interaction
landscapes for four classes of rigid $\alpha$-polyalanine helices. The
symmetry-adapted representations reproduce the equilibrium well-depth and
separation fields with errors of only a few meV and approximately
$6$--$8$~m\AA{}, respectively, while representing more than \num{300000}
reference energy values per class with tens to a few hundred Fourier
coefficients. For the full angle-dependent-$\alpha$ representation, the
interaction-energy RMSE is $3.26$--$3.93$~meV within $2k_{\mathrm B}T$ and
$4.36$--$4.82$~meV within $4k_{\mathrm B}T$ at $300$~K, remaining well below
the thermal energy scale in the vicinity of the pairwise minima. Further
pruning reduces the number of retained coefficients with only small changes
in reconstruction accuracy. In the full representation,
$D(\chi,\psi)$, $r_e(\chi,\psi)$, and $\alpha(\chi,\psi)$ together
describe the orientational dependence of the interaction strength,
equilibrium separation, and radial shape. The constant-$\alpha$ form
provides a more compact reduced representation when near-equilibrium accuracy
is sufficient.

The resulting analytical potentials can be evaluated directly without
multidimensional table interpolation and can be used in molecular dynamics.
Microcanonical tests showed stable total-energy conservation, and
proof-of-concept annealing simulations produced low-temperature configurations
with local ordering tendencies qualitatively consistent with motifs
identified previously by Monte Carlo simulated annealing. The
parameterization and model-reduction workflow is implemented in the
open-source \texttt{ChiMorse} package~\autocite{chimorse}.

The present formulation is restricted to rigid particles described by an
intermolecular separation and two periodic orientational coordinates, with
additional configurational degrees of freedom either fixed or treated
separately. Extensions to axial displacement, molecular tilt, internal
conformational changes, and more general anisotropic interaction spaces
provide natural directions for future work, together with systematic studies
of finite-temperature assembly and kinetics using the analytical potentials.

More broadly, the direct correspondence between Fourier modes and
orientational periodicities provides a natural basis for constructing or
modifying model interactions with prescribed symmetries. Symmetry constraints
imposed at the level of the Fourier basis remain exact even when the sampled
interaction landscape cannot be reproduced with arbitrary quantitative
accuracy, including in off-equilibrium regions. The representation therefore
separates controlled, interpretable symmetry content from the numerical
accuracy associated with a finite harmonic expansion. Although inverse
construction is not pursued quantitatively here, this feature provides a
natural starting point for future work on anisotropic interactions with
prescribed orientational characteristics and their consequences for
self-assembled structures.


\section*{Supplementary Material}

See the supplementary material for details of the reference-data sampling and
symmetry completion, interchange-symmetry derivations, additional angular
parameter landscapes and convergence analyses, full-model coefficient
pruning, reconstruction errors of the final pruned representations, and
molecular-dynamics simulation and energy-conservation tests.

\section*{Acknowledgements}
Authors acknowledge funding by German Research Foundation (DFG), TRR-386, TP A4 and B2, project number 514664767.

\printbibliography

@article{Hadis_2026,
author = {Ghodrati, Hadis and Preis, Kevin and Ha Nguyen, Thi Ngoc and Tegenkamp, Christoph and Gemming, Sibylle and Kelling, Jeffrey and G{\"u}nther, Florian},
title = {Simulation of Self-Assembled Monolayers of Polyalanine $\alpha$-Helices: Development and Application of an Effective Potential for Film Structure Predictions},
journal = {ACS Applied Materials \& Interfaces},
volume = {18},
number = {21},
pages = {30467-30479},
year = {2026},
doi = {10.1021/acsami.6c01087},
note ={PMID: 42172137},
}

@MISC{chimorse,
  title     = {chimorse python package},
  author    = {Ghodrati, Hadis and Kelling, Jeffrey},
  publisher = {Zenodo},
  doi = {10.5281/zenodo.22071766},
  year =  {2026},
}

@MISC{data_PA,
  title     = {Interaction potential of parallel and antiparallel aligned, infinite polyalanine alpha-helices as obtained from SCC-DFTB + UFF calculations},
  author    = {G{\"u}nther, Florian Steffen and Preis, Kevin and Gemming, Sibylle and Kelling, Jeffrey and Ghodrati, Hadis},
  publisher = {Zenodo},
  doi = {10.5281/zenodo.21904447},
  year =  {2026},
}

@article{inverse_SAM_2D_2006,
  title = {Designed interaction potentials via inverse methods for self-assembly},
  author = {Rechtsman, Mikael and Stillinger, Frank and Torquato, Salvatore},
  journal = {Phys. Rev. E},
  volume = {73},
  pages = {011406},
  numpages = {12},
  year = {2006},
  publisher = {American Physical Society},
  doi = {10.1103/PhysRevE.73.011406},
}

@article{inverse_SAM_2018,
    author = {Adorf, Carl S. and Antonaglia, James and Dshemuchadse, Julia and Glotzer, Sharon C.},
    title = {Inverse design of simple pair potentials for the self-assembly of complex structures},
    journal = {The Journal of Chemical Physics},
    volume = {149},
    number = {20},
    pages = {204102},
    year = {2018},
    doi = {10.1063/1.5063802},
}

@article{SAM_simulation_review_2016,
author = {Wen, Jin and Li, Wei and Chen, Shuang and Ma, Jing},
year = {2016},
pages = {22757-22771},
title = {Simulations of molecular self-assembled monolayers on surfaces: packing structures, formation processes and functions tuned by intermolecular and interfacial interactions},
volume = {18},
journal = {Physical Chemistry Chemical Physics},
doi = {10.1039/C6CP01049K}
}

@article{SAM_KMC_2007,
    author = {Haran, Mohit and Goose, Joseph E. and Clote, Nicolas P. and Clancy, Paulette},
    title = {Multiscale Modeling of Self-Assembled Monolayers of Thiophenes on
Electronic Material Surfaces},
    journal = {Langmuir},
    volume = {23},
    number = {9},
    pages = {4897-4909},
    year = {2007},
    issn = {0743-7463},
    doi = {10.1021/la063059d}
}

@article{SAM_structure_defects_2013,
    author = {Claridge, Shelley A. and Liao, Wei-Ssu and Thomas, John C. and Zhao, Yuxi and Cao, Huan H. and Cheunkar, Sarawut and Serino, Andrew C. and Andrews, Anne M. and Weiss, Paul S.},
    title = {From the bottom up: dimensional control and characterization in molecular monolayers},
    journal = {Chemical Society Reviews},
    volume = {42},
    number = {7},
    pages = {2725-2745},
    year = {2013},
    issn = {0306-0012},
    doi = {10.1039/c2cs35365b},
}

@article{SAM_dynamics_defects_2005,
    author = {Vericat, C. and Vela, M. E. and Salvarezza, R. C.},
    title = {Self-assembled monolayers of alkanethiols on Au(111): surface structures, defects and dynamics},
    journal = {Physical Chemistry Chemical Physics},
    volume = {7},
    number = {18},
    pages = {3258-3268},
    year = {2005},
    issn = {1463-9076},
    doi = {10.1039/b505903h},
}

@article{SAM_organic_e_2017,
    author = {Casalini, Stefano and Bortolotti, Carlo Augusto and Leonardi, Francesca and Biscarini, Fabio},
    title = {Self-assembled monolayers in organic electronics},
    journal = {Chemical Society Reviews},
    volume = {46},
    number = {1},
    pages = {40-71},
    year = {2017},
    issn = {0306-0012},
    doi = {10.1039/c6cs00509h},
}

@article{bio_helices_pot_2007,
  title = {Structure and interactions of biological helices},
  author = {Kornyshev, Alexei A. and Lee, Dominic J. and Leikin, Sergey and Wynveen, Aaron},
  journal = {Rev. Mod. Phys.},
  volume = {79},
  pages = {943--996},
  numpages = {0},
  year = {2007},
  publisher = {American Physical Society},
  doi = {10.1103/RevModPhys.79.943},
}

@article{spintronic_review_2016,
title = {A review on organic spintronic materials and devices: II. Magnetoresistance in organic spin valves and spin organic light emitting diodes},
journal = {Journal of Science: Advanced Materials and Devices},
volume = {1},
number = {3},
pages = {256-272},
year = {2016},
issn = {2468-2179},
doi = {10.1016/j.jsamd.2016.08.006},
author = {Rugang Geng and Hoang Mai Luong and Timothy Tyler Daugherty and Lawrence Hornak and Tho Duc Nguyen},
}

@article{CISS_review_2025,
author ="Mishra, Suryakant and Jones, Andrew C. and Fontanesi, Claudio",
title  ="Recent advancements in chiral spintronics: from molecular-level insights to device applications. A prospect based on the interplay between physical and chemical properties of chiral systems",
journal  ="J. Mater. Chem. C",
year  ="2025",
volume  ="13",
pages  ="2121-2134",
publisher  ="The Royal Society of Chemistry",
doi  ="10.1039/D4TC03453H"}

@article{CISS_SAM_experiment_2024,
  title     = "Chirality-induced magnet-free spin generation in a semiconductor",
  author    = "Liu, Tianhan and Adhikari, Yuwaraj and Wang, Hailong and Jiang,
               Yiyang and Hua, Zhenqi and Liu, Haoyang and Schlottmann, Pedro
               and Gao, Hanwei and Weiss, Paul S and Yan, Binghai and Zhao,
               Jianhua and Xiong, Peng",
  journal   = "Adv. Mater.",
  publisher = "Wiley",
  volume    =  36,
  number    =  36,
  pages     = "e2406347",
  year      =  2024,
  doi = "10.1002/adma.202406347",
  language  = "en"
}

@article{book_computational_2022,
    
author={Reis, Heribert  and Żuchowski, Piotr  and Grubisic, Sonja },
           
title={Editorial: Computational Methods for the Description of Intermolecular Interactions and Molecular Motion in Confining Environments},
journal={Frontiers in Chemistry},
volume = {10},
pages = {941269},
year={2022},
doi={10.3389/fchem.2022.941269},  
}

@article{comp_interaction_2025,
    author = {Mieres-Perez, Joel and Almeida-Hernandez, Yasser and Sander, Wolfram and Sanchez-Garcia, Elsa},
    title = {A Computational
Perspective to Intermolecular Interactions
and the Role of the Solvent on Regulating Protein Properties},
    journal = {Chemical Reviews},
    volume = {125},
    number = {15},
    pages = {7023-7056},
    year = {2025},
    doi = {10.1021/acs.chemrev.4c00807},
}

@article{Sutherland_chiralCG_3D_2019,
author ={Sutherland, B. J. and Olesen, S. W. and Kusumaatmaja, H. and Morgan, J. W. R. and Wales, D. J.},
title  ={Morphological analysis of chiral rod clusters from a coarse-grained single-site chiral potential},
journal  ={Soft Matter},
year  ={2019},
volume  ={15},
pages  ={8147-8155},
publisher  ={The Royal Society of Chemistry},
doi  ={10.1039/C9SM01343A}}

@article{Curco_CG_oriented_helical_2007,
author = {Curcó, David and Nussinov, Ruth and Alemán, Carlos},
title = {Coarse-graining the Self-assembly of $\beta$-helical Protein Building Blocks},
journal = {The Journal of Physical Chemistry B},
volume = {111},
number = {50},
pages = {14006-14011},
year = {2007},
doi = {10.1021/jp075386f},
note ={PMID: 18027921}
}

@article{babadi_CG_2006,
    author = {Babadi, M. and Everaers, R. and Ejtehadi, M. R.},
    title = {Coarse-grained interaction potentials for anisotropic molecules},
    journal = {The Journal of Chemical Physics},
    volume = {124},
    number = {17},
    pages = {174708},
    year = {2006},
    issn = {0021-9606},
    doi = {10.1063/1.2179075},
}

@article{Noid_CG_review_2023,
author = {Noid, W. G.},
title = {Perspective: Advances, Challenges, and Insight for Predictive Coarse-Grained Models},
journal = {The Journal of Physical Chemistry B},
volume = {127},
number = {19},
pages = {4174-4207},
year = {2023},
doi = {10.1021/acs.jpcb.2c08731},
    note ={PMID: 37149781}

}

@article{ML_coarse_grained_MD_2019,
author = {Wang, Jiang and Olsson, Simon and Wehmeyer, Christoph and P{\'e}rez, Adrià and Charron, Nicholas E. and de Fabritiis, Gianni and No{\'e}, Frank and Clementi, Cecilia},
title = {Machine Learning of Coarse-Grained Molecular Dynamics Force Fields},
journal = {ACS Central Science},
volume = {5},
number = {5},
pages = {755-767},
year = {2019},
doi = {10.1021/acscentsci.8b00913},
    note ={PMID: 31139712}
}

@article{ML_CG_2023,
  title     = "Machine learning coarse-grained potentials of protein
               thermodynamics",
  author    = "Majewski, Maciej and P{\'e}rez, Adri{\`a} and Th{\"o}lke,
               Philipp and Doerr, Stefan and Charron, Nicholas E and Giorgino,
               Toni and Husic, Brooke E and Clementi, Cecilia and No{\'e},
               Frank and De Fabritiis, Gianni",
  journal   = "Nat. Commun.",
  publisher = "Springer Science and Business Media LLC",
  volume    =  14,
  number    =  1,
  pages     = "5739",
  year      =  2023,
  copyright = "https://creativecommons.org/licenses/by/4.0",
  language  = "en",
  doi       = "10.1038/s41467-023-41343-1"
}

@article{Morse_original,
  title = {Diatomic Molecules According to the Wave Mechanics. II. Vibrational Levels},
  author = {Morse, Philip M.},
  journal = {Phys. Rev.},
  volume = {34},
  pages = {57--64},
  numpages = {0},
  year = {1929},
  publisher = {American Physical Society},
  doi = {10.1103/PhysRev.34.57},
}

@article{LJ_classic,
 ISSN = {09501207},
 URL = {http://www.jstor.org/stable/94265},
 author = {J. E. Jones},
 journal = {Proceedings of the Royal Society of London. Series A, Containing Papers of a Mathematical and Physical Character},
 number = {738},
 pages = {463--477},
 publisher = {The Royal Society},
 title = {On the Determination of Molecular Fields. II. From the Equation of State of a Gas},
 volume = {106},
 year = {1924}
}

@article{GayBerne_1995,
title = {A generalized Gay-Berne intermolecular potential for biaxial particles},
journal = {Chemical Physics Letters},
volume = {236},
number = {4},
pages = {462-468},
year = {1995},
issn = {0009-2614},
doi = {10.1016/0009-2614(95)00212-M},
author = {R. Berardi and C. Fava and C. Zannoni}}

@article{helical_potential_1997,
author = {Kornyshev, A. A. and Leikin, S.},
title = {Theory of interaction between helical molecules},
journal = {The Journal of Chemical Physics},
volume = {107},
number = {9},
pages = {3656-3674},
year = {1997},
issn = {0021-9606},
doi = {10.1063/1.475320},
}

@article{patchy_review_2011,
author ="Bianchi, Emanuela and Blaak, Ronald and Likos, Christos N.",
title  ="Patchy colloids: state of the art and perspectives",
journal  ="Phys. Chem. Chem. Phys.",
year  ="2011",
volume  ="13",
pages  ="6397-6410",
publisher  ="The Royal Society of Chemistry",
doi  ="10.1039/C0CP02296A"}

@article{torsional_fourier_1989,
author = {Chung-Phillips, Alice},
title = {Methods for the Fourier-series expansion of torsional energies},
journal = {Journal of Computational Chemistry},
volume = {10},
number = {5},
pages = {733-747},
doi = {10.1002/jcc.540100514 },
year = {1989}
}
\end{document}